\documentclass[intlimits,twoside,a4paper]{article}
\usepackage{wrapfig}
\usepackage{amssymb}
\usepackage[T2A]{fontenc}
\usepackage[cp1251]{inputenc}
\usepackage{amsmath}
\usepackage{graphicx}

\usepackage{cmpj2}

\issue{2015}{18}{1}{13603}
\doinumber{10.5488/CMP.18.13603}

\title[Density Anomaly of the SPC/E water model. Molecular Dynamics.]
{Temperature dependence of the microscopic structure and density anomaly
of the SPC/E and TIP4P-Ew water models. Molecular dynamics simulation results}

\author[E. Galicia-Andr\'{e}s, H. Dominguez, O. Pizio]
{E. Galicia-Andr\'{e}s\refaddr{label1}, H. Dominguez\refaddr{label2},
O. Pizio\refaddr{label1}\thanks{E-mail: pizio@unam.mx}}

\addresses{
\addr{label1}Instituto de Qu\'{i}mica, Universidad Nacional Aut\'{o}noma de M\'{e}xico,
Circuito Exterior, 04510, M\'{e}xico, D.F., M\'{e}xico
\addr{label2}Instituto de Investigaciones en Materiales,
Universidad Nacional Aut\'{o}noma de M\'{e}xico,
Circuito Exterior, 04510, M\'{e}xico, D.F., M\'{e}xico}

\date{Received November 6, 2014, in final form January 13, 2015}

\begin{document}
\maketitle

\begin{abstract}
We have investigated temperature trends of the microscopic structure
of the SPC/E  and TIP4P-Ew water models in terms of the pair distribution
functions, coordination numbers,
the average number of hydrogen bonds, the distribution of bonding states
of a single molecule as well as the angular distribution of molecules
by using the constant pressure molecular dynamics simulations.
The evolution of the structure is put in correspondence with the dependence of
water density on high temperatures down to the region of temperatures
where the system becomes supercooled. It is shown that the fraction of molecules with
three and four bonds determine the maximum density for both models.
Moreover, the temperature dependence of the dielectric constant is obtained and analyzed.


\pacs 61.20.-p, 61.20.Gy, 61.20.Ja, 65.20.Jk

\end{abstract}

\section{Introduction}

It is our honor and pleasure to contribute to this special issue and to
dedicate this paper to Prof. Dr.~Douglas Henderson on the occasion of his
$80^\textrm{th}$ birthday.
Almost four decades ago, Barker and Henderson published their seminal paper
``What is ``liquid''? Understanding the states of matter'', that comprehensively describes
basic theoretical methods to explain the equilibrium properties of simple liquids \cite{doug1}. These
methods have made substantial progress since that time. In particular, computer simulations techniques
have become essentially powerful and common tools in the theory of liquids, mixtures and solutions.
Computer simulations permitted to reach enormous progress in understanding the  liquids
with hydrogen bonds, more specifically, water and aqueous solutions, for needs of
physical chemistry and for several inter-disciplinary areas involving complex biological
molecules and membranes. In particular, Henderson with co-workers was actively interested
in the properties and behavior of inhomogeneous aqueous solution, see e.g. \cite{doug2,doug3,doug4}.

The studies of water have long history that has been documented in an enormous number
of publications. Of particular interest are the peculiar thermodynamic and dynamic
properties of water manifested in a set of anomalies. In spite of formal equivalence of
the relation between the microscopic structure and thermodynamic properties discussed
in great detail by Barker and Henderson for simple homogeneous liquids \cite{doug1},
in the case of water it is inevitable to involve the notion of hydrogen bonding.
Many works have dealt with the exploration of hydrogen bonds network structure,
its dynamics and life-time of bonds at an early stage of computer simulations
of aqueous systems, see e.g. a few original papers \cite{rahman,jorgensen} as well
as later reviews \cite{lang,guillot,brovchenko1}, that contain an extensive list of
references on the subject. Moreover, it is known that unique properties of water
specifically manifest themselves in the region of states below the freezing point
corresponding to the supercooled liquid, see e.g. a recent review \cite{holten}
and references therein.

It is seducing to relate at least some of water anomalies with the transformations
occuring in the hydrogen bonds cooperative structure. Therefore,
in this short report our focus is in the description of the evolution of the average number of
H-bonds per water molecule and in the distribution of the numbers of H-bonds
per molecule upon temperature changes. Moreover, we would like to relate
the trends of the behavior of these properties with the density anomaly. While the
properties in  question describe the formation of ``agglomerates'' of bonded
molecules at relatively short-range distances between them, we would like to get additional
insights into the simultaneously occuring changes of the long-range correlations between water
molecules in terms of the dielectric constant.

One of the most studied anomalies of water is its nonmonotonous temperature
dependence of density, $\rho(T)$, see e.g. a review providing a nice historical
insight \cite{brovchenko1}.
Several non-polarizable water models with different number of interaction sites
are capable of reproducing the presence of maximum density of water
with different precision.

Namely, the temperature of maximum density (TMD) at ambient
pressure for most frequently used models is observed at 253~K (TIP4P),
at 278~K (TIP4P/2005), at 274.15~K (TIP4P-EW), at 241~K (SPC/E) and at 182~K (TIP3P).
These values for temperature are taken from the table~2 of the review
by Vega and Abascal \cite{vega1}. The experimental results for these and other properties
used by the authors are taken from \cite{saul}.
The experiment yields 277~K for the TMD of real water.
On the other hand, the water density at TMD is 0.988 (TIP4P), 0.993 (TIP4P/2005),
0.994 (SPC/E), 0.98 (TIP3P), while the experimental value is 0.997.
Some deviation from these values reported in different works result
from technical details of each simulation.

The origin of density anomaly is commonly attributed to the evolution of
hydrogen bond network with the formation of an open structure as the temperature is lowered
 or, on the other hand,  manifests itself as the reflection of a possible
liquid-liquid phase transition observed by simulations in some water models \cite{brovchenko1}.
The latter and other scenarios for peculiar properties of supercooled water were
documented in~\cite{holten}.

To summarize this brief introduction, our intention in this work is to
to present our own data of molecular dynamics simulations of the temperature dependence of
water density, $\rho(T)$, in an ample interval of temperature (from 370~K down to 170~K) at ambient pressure
by using the SPC/E model. Next, we would like to compare the obtained data with the results  previously
reported by other authors. However, the principal focus is to follow
changes of water structure in terms of different descriptors involving both the
hydrogen bonds and the $\rho(T)$ dependence
in order to establish a clear relationship between them.
Nevertheless, for the sake of comparison of our observations for the SPC/E, and to make the
conclusions more sound, similar results for the TIP4P-Ew water model \cite{horn} are presented as well.
Our interest in the SPC/E model is motivated by its wide use for different purposes,
in particular for the description of thermodynamic properties of mixtures.
Extending the previous research in mixtures, having in  mind a more ample project presently being developed
in this laboratory,
our intention  is to use the SPC/E model as a starting point in exploring water--alcohol
and water--DMSO solutions by using molecular dynamics simulations.

\section{Simulation details}

All our simulations of water models were performed in the isothermal-isobaric ensemble at
ambient pressure. The initial configuration was constructed with 2000 water molecules
placed in a cubic array in a simulation box. Then, the system was simulated
in an ample range of temperatures, from a high temperature 370~K down to 180~K.
The DL-POLY Classic package was employed \cite{forest}. We used the Berendsen thermostat and
barostat, the running timestep was set to 0.002~ps.  Commonly, the periodic boundary
conditions were used. The short range interactions were cut-off at 11~{\AA},
whereas the long-range interactions were handled by the
Ewald method with the precision $10^{-4}$.
After equilibration, several sets of simulation runs, each for 6--8~ns, with restart
from the previous configuration, were performed to obtain averages for data analysis.
The overall length of simulation was dictated by the stability of internal energy, density and
the dielectric constant, all being the functions of time.

\section{Results and discussion}

\begin{figure}[!b]
\begin{center}
\includegraphics[width=0.49\textwidth,clip]{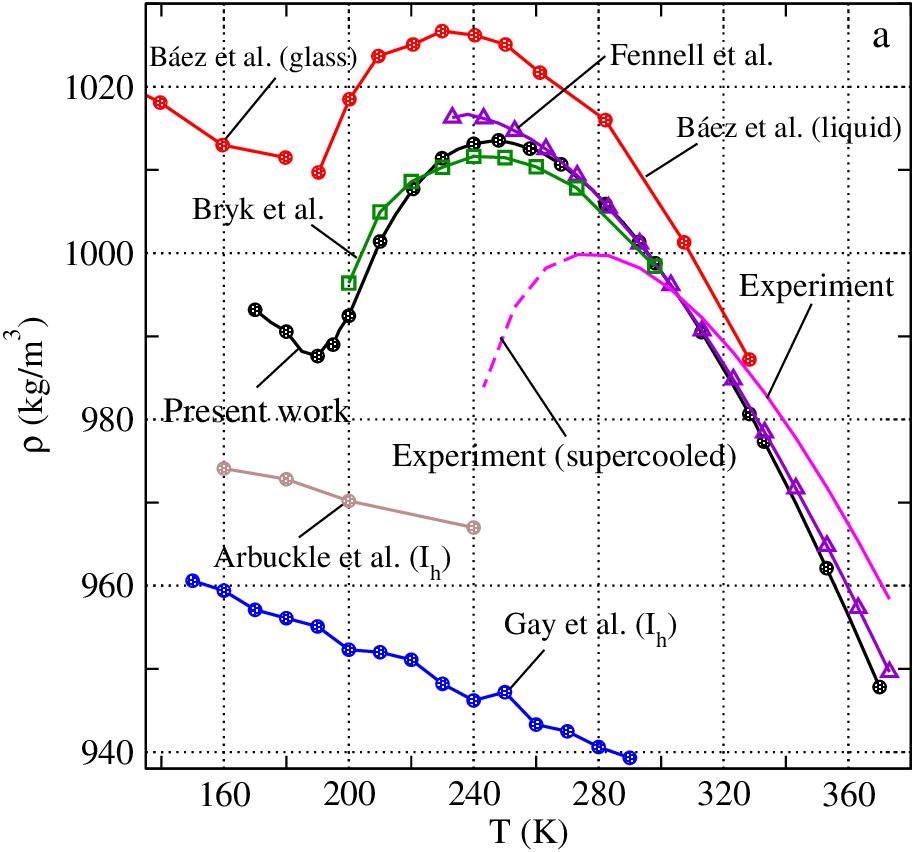}
\hfill
\includegraphics[width=0.505\textwidth,clip]{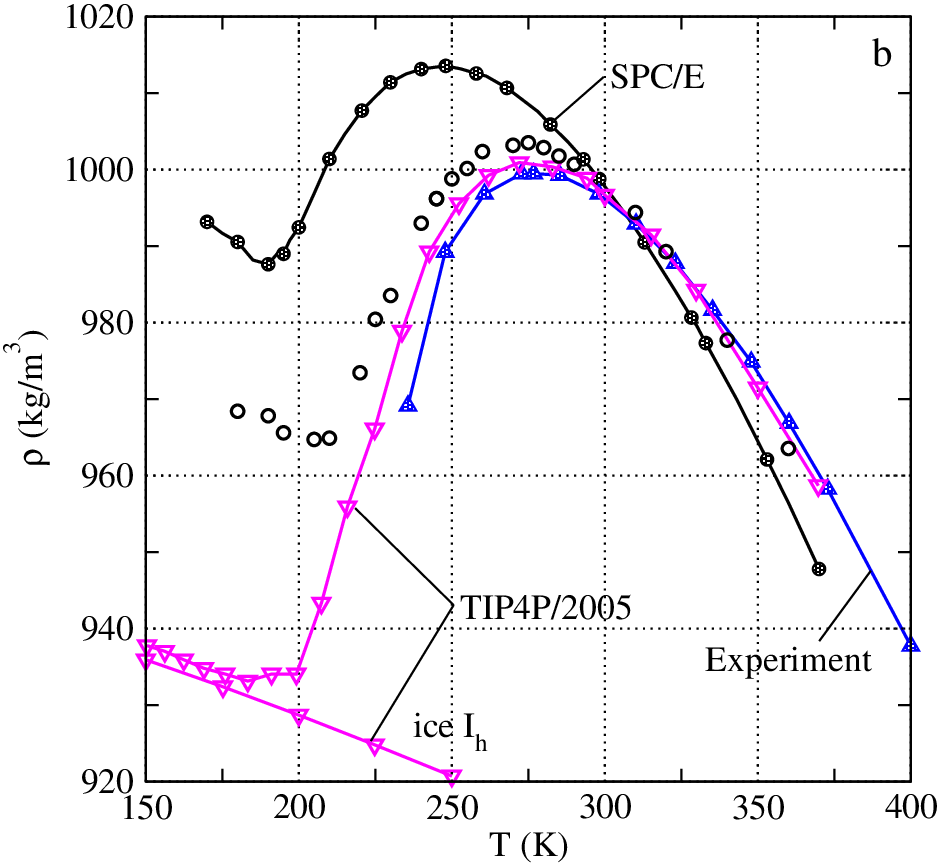}
\end{center}
\caption{(Color online) Panel a: Temperature dependence of the SPC/E model water density
in the liquid and glassy state as well as of $I_\textrm{h}$ ice at a constant ambient pressure.
The results of different authors are marked in the figure. Panel b: A comparison
of the density dependence of the SPC/E and TIP4P type models. Circles show our
results for the TIP4P-Ew model.}\label{fig1}
\protect
\end{figure}

The temperature dependence of the density of water and the stability of
several ice phases in the framework of the SPC/E model were studied in several works,
see e.g. \cite{haymet1,haymet2,haymet3,clancy1,clancy2,clancy3,dill}.
In figure~\ref{fig1}~(a)  we show the results obtained by different
authors for the SPC/E water and the results of the present work.
The liquid and glass branches, denominated in figure~\ref{fig1}~(a) as Baez et al., were constructed
using the points reported in
table~II and in table~III of reference \cite{clancy1}. The data follow from the simulations of
360 molecules. It was assumed that the cut-off distance is at 9~{\AA} and the effect
of interactions at larger distances was neglected. The density maximum ($\rho=1.0265$~g/cm$^3$)
was obtained at 235~K. In the later work from that laboratory, the density
values for disordered hexagonal ice ($I_\textrm{h}$) in the interval between 170~K and 220~K, were
reported, see reference \cite{clancy3}. These data are also shown in figure~\ref{fig1}~(a).
However, it is important to mention that the results refer to
the modified SPC/E model, i.e., by multiplying the interaction potential by a switching function
between 6~{\AA} and 9~{\AA} to make the forces and the potential continuous at a cut-off distance.
The modification of the potential and the application of Ewald summation technique to include
the long-range interactions also resulted in the changes of the estimates for TMD,
in \cite{clancy2,clancy3} the improved values are $\rho=1.003$~g/cm$^3$ and $T\approx 260$~K.
The algorithmic peculiarities of these series of works include a profound exploration of the
effects of Ewald summation on the temperature dependence of density and other properties
of the system as well as calculations of the Gibbs free energy of distinct phases which
requires thermodynamic integration.
The line describing the density of stable $I_\textrm{h}$ ice obtained by Gay et al. is
reproduced from \cite{haymet1}. It was mentioned therein that the stability of this specific
crystalline phase crucially depends on the number of molecules in the simulation cell.
It is worth mentioning that the authors performed simulations in isostress, constant temperature
ensemble and estimated the melting point by analyzing the time evolution of the properties
of the system upon heating of the solid phase.

Quite recently, Vega et al. \cite{vega2} evaluated the
melting temperature of the most popular non-polarizable  water models.
For the SPC/E model, the melting temperature value of ice $I_\textrm{h}$ and the
densities of liquid water and ice are shown in figure~\ref{fig1}~(a).
Our data for the liquid density of the model starting from a rather high temperature down to the region
of supercooled liquid are compared with the previously published results by Bryk and Haymet
\cite{haymet3} and with very recent results by Fennell et al. \cite{dill}.
A small difference between the curve obtained in \cite{haymet3}
and the present results in the entire interval of temperatures is presumably caused
by technical details, namely by the distinct cut-off value. On the other hand, it is worth mentioning
that Fennell et al. \cite{dill} slightly modified the original SPC/E model by fitting the
dielectric constant value at room temperature to the experimental result.
Our data locate the TMD at 245~K, i.e., in between the commonly used value 241~K,
see e.g. \cite{vega1,haymet3}, and the result for the optimized model by Fennell et al. 247~K.
As for the value of density at TMD, our result is quite close to the data
reported in~\cite{haymet3}.

In general terms, the water density values in relation to temperature grow
in the interval that starts at high temperatures and is delimited below by the TMD.
Next, if the temperature
decreases further, the density decreases until it reaches a minimum value, $T_\textrm{min}$,
our data locate the minimum at $T=190$~K. The minimum value of density was observed for
a set of water models, e.g., for ST2 \cite{brovchenko1} and TIP4P-2005 \cite{vega3}.
In the case of the SPC/E model,
the melting temperature $T_\textrm{m}$ is located in the interval between the TMD and $T_\textrm{min}$,
and the region between the melting point and $T_\textrm{min}$ actually corresponds to the supercooled
states. Finally, if the temperature decreases below the $T_\textrm{min}$, the density grows again
as in a normal fluid, but this growth is not as rapid as in the interval of above the TMD.

In our opinion, a plausible explanation of the presence of the density minimum that involves
structural changes in water was given in \cite{brovchenko1}. It refers to the behavior of a simple
liquid that shrinks upon cooling due to the increasing effects of attraction between particles
with a decreasing temperature. However, in a certain temperature interval, this normal
behavior alters due to augmenting effects of hydrogen bonding between molecules and the resulting
formation of ``complexes'' or ``agglomerates''. Cooperative behavior of these entities result
in the growth of an expanded, open structure. The effects of directional hydrogen bonding
overcome the effects of an attractive interaction. Consequently, the fluid  density decreases
in a certain interval of temperatures, evidently specific for each model. However, the directional
interactions are characterized by saturation. Therefore, upon further cooling, when changes
of the structural units become weaker, compared with the previous regime (as documented for example by
the changes of the average number of hydrogen bonds per molecule), then the fluid density
grows again  similar to the temperature interval above the TMD.

However, the above discussion  is incomplete and some thermodynamic arguments must be taken into
account. The liquid is the stable equilibrium phase for temperatures above the melting temperature and
it is unstable below a stability temperature, whereas in between, the liquid is metastable with
respect to crystal ice \cite{chandler}. In general, metastable liquids and specifically
water, can survive even at temperatures well below the melting point until homogeneous
nucleation takes place and water converts into ice \cite{vega4}. Therefore, a certain
part of the curve $\rho(T)$ presented in figure~\ref{fig1}~(a) for the SPC/E model (below the melting point)
characterized by a descending density can correspond to metastable, ``long-life'' states
(on the simulation time scale). Similar comments concern the behavior of the TIP4P-2005 model, see
figure~4 of reference \cite{vega3} and the ST2 model \cite{brovchenko1,chandler}.

Due to finite-size effects and consequent details in implementation of Ewald sums as well as due to
finite sampling time, a possible spontaneous crystallization, however, is not observed in
many works for a set of water models, as well as in our present simulations.
Debates concerning this problem just started very recently, see e.g. \cite{chandler,vega4}.
These authors provide a comprehensive discussion of various issues relevant to the problem
using their own data coming from the Monte Carlo and molecular dynamics simulations and analyzing
the previously published results for different water models from the literature.
Some other papers from the special issue of the Journal of Chemical Physics devoted to
confined and interfacial water published in November 2014 can be useful for the reader as well.

Our final comments concerning a set of curves shown in figure~\ref{fig1}~(a) refer to the glassy water
in the framework of the SPC/E model. Actually, water, if considered below the homogeneous nucleation
temperature, should be in a glassy form, see e.g., a rather comprehensive discussion of the issue
in \cite{stanley}. In the case of the SPC/E model, the glass transition temperature was located
around 177~K \cite{clancy1}. At this temperature, the dependence of internal energy on temperature
shows a kink (we reproduced the results of that work by performing long simulations
with 2000 molecules and located the kink around $T\approx185$~K). Thus, the augmenting density region at low
temperatures refers in part to the amorphous states that cannot be distinguished in terms of
the structural characteristics such as the pair distribution functions coming from the
present simulations.

In figure~\ref{fig1}~(b), we present the temperature dependence of density of the SPC/E model and compare it
with our results obtained for the TIP4P-Ew model performed in the framework of the above described
methodology. In addition, the results by H.~Pi et al. \cite{vega3} for the liquid and ice $I_\textrm{h}$
phases are given as well. The TIP4P-Ew model reproduces the experimental result for the TMD
and an overall dependence of density at ambient pressure \cite{handbook}
much better compared to the SPC/E.
All the models involved in this figure (SPC/E, TIP4P-Ew, TIP4P-2005) predict the existence of the
density minimum and augmenting density at lower temperatures.
However, it is worth mentioning that the density of ice $I_\textrm{h}$ substantially differs from the
density of supercooled water (and of its glassy form) for the SPC/E model [see figure~\ref{fig1}~(a)], whereas
in the case of TIP4P-2005 model, the liquid phase density and that of ice are close to
each other as it follows from the simulations by H. Pi et al. \cite{vega3}, see figure~\ref{fig1}~(b).
Similar behavior concerning the density dependences has been documented recently
by J.~Alejandre et al. \cite{alejandre} for the model from the TIP4P family developed by
fitting the dielectric constant, the maximum density and the equation of state at low
pressure.
Unfortunately, we are unaware of the simulation results for the $I_\textrm{h}$ ice phase of
TIP4P-Ew model at room pressure to make our discussion more general at this point.
However, it seems that the differences between liquid and ice phase densities in the
region of low temperatures are very sensitive to the minor changes of parameters
describing different models.

The microscopic structure of a set of water models coming out from MD simulations
with respect to the experimental diffraction data at room temperature and ambient pressure
was discussed in great detail by Pusztai et al. \cite{pusztai1} by using the
protocol \cite{pusztai2} combining the experimental total scattering structure factor
and partial radial distribution functions from simulations. It was shown that the TIP4P-2005
model exhibits the best consistency with the experimental structure compared to other
commonly used models. Nevertheless, the SPC/E, ST2 and TIP4P results are not seriously inconsistent
at ambient conditions. Unfortunately, such an analysis of the structure of water has not
been performed for other temperatures and pressures due to experimental difficulties.

\begin{figure}[!b]
\begin{center}
\includegraphics[width=0.49\textwidth,clip]{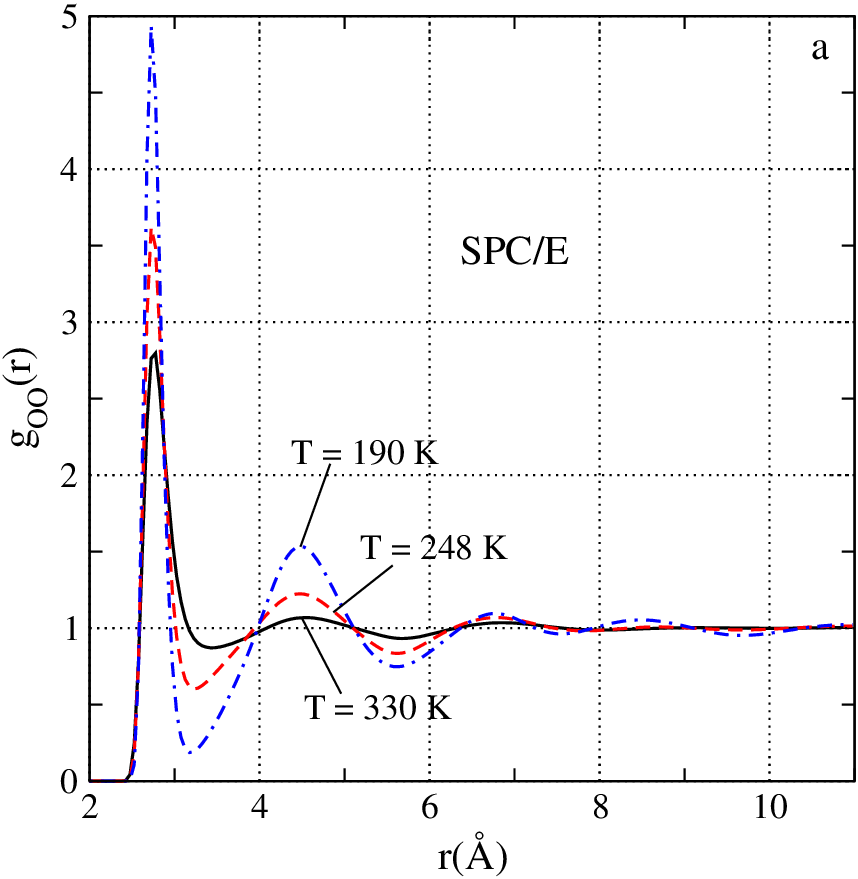}
\hfill
\includegraphics[width=0.505\textwidth,clip]{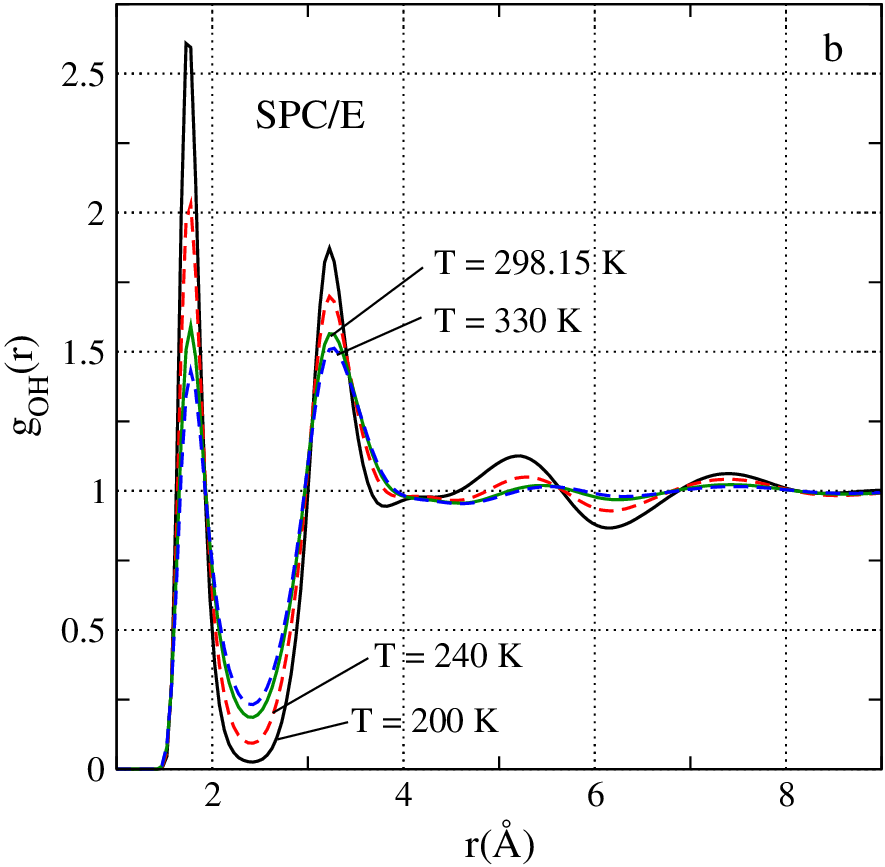}
\end{center}
\caption{(Color online) Oxygen--oxygen (panel a) and oxygen--hydrogen
(panel b) pair distribution functions of the SPC/E model at different temperatures.}\label{fig2}
\protect
\end{figure}

A few examples of the radial distribution functions, $g_{ij}$, where $i,j$
stand for O and H, obtained in our NPT simulations of the SPC/E model are presented in two
panels of figure~\ref{fig2}.
It can be seen that the height of the first and the subsequent maxima of both functions, $g_\textrm{OO}(r)$
and $g_\textrm{OH}$, increases with a decreasing temperature in the entire interval of temperatures studied.
These trends are well pronounced upon cooling in the range starting from a high temperature down
to around 190~K, further. At even lower temperatures, the overall shape of $g_{ij}(r)$ is
similar to that observed at 190~K. However, changes of the structure are much less pronounced
if one moves to 180~K and 170~K. We omit these results for the sake of brevity.
The functions $g_\textrm{OO}(r)$ and $g_\textrm{OH}(r)$ are given mostly as an illustration, because their
characteristics are necessary to introduce the concept of hydrogen bonds between molecules.

\begin{figure}[!t]
\begin{center}
\includegraphics[width=0.48\textwidth,clip]{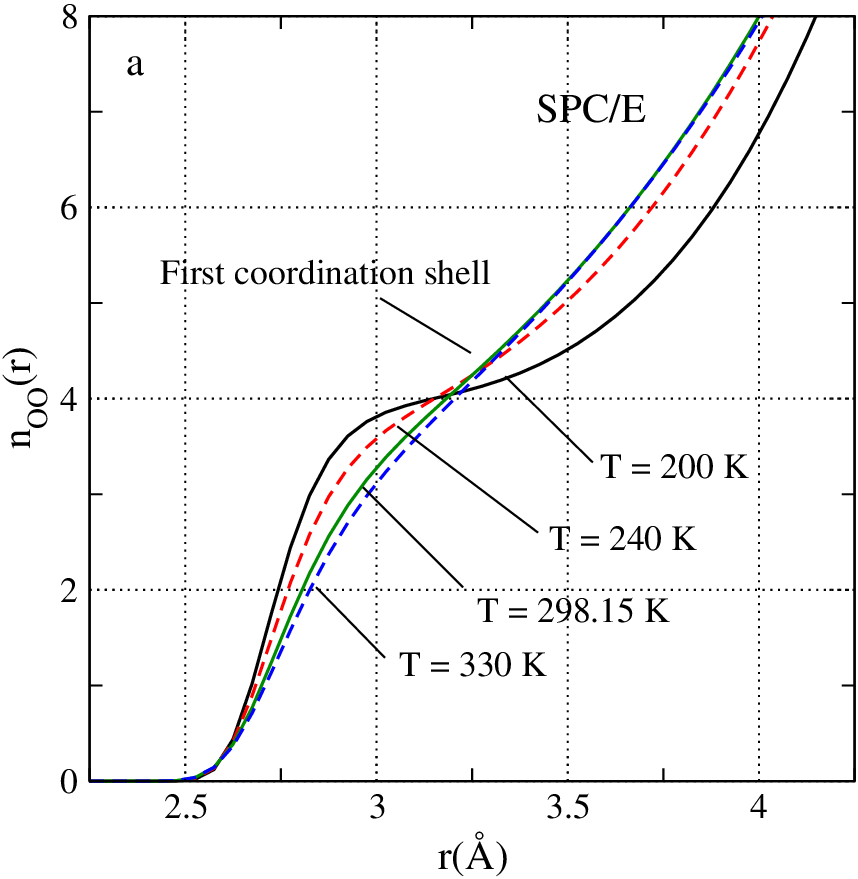}
\hfill
\includegraphics[width=0.50\textwidth,clip]{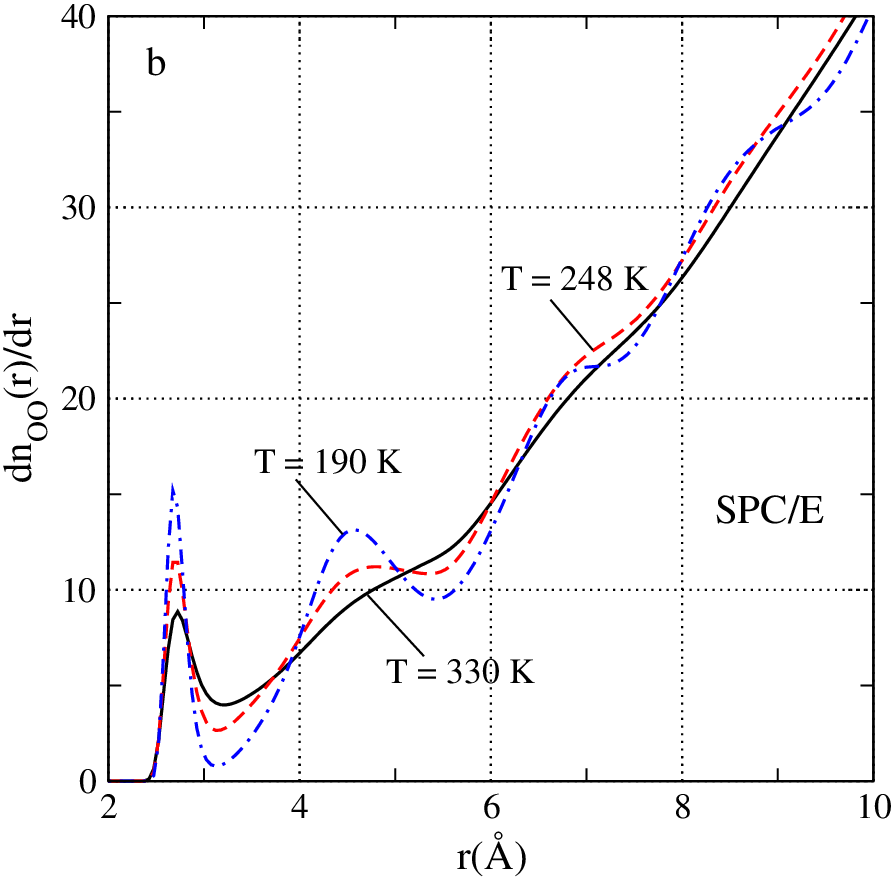}
\end{center}
\caption{(Color online) Oxygen--oxygen coordination number of the SPC/E model
at different temperatures (panel a). The derivative of the oxygen--oxygen coordination number
at different temperatures (panel b).}\label{fig3}
\protect
\end{figure}

In terms of the coordination number, $n_\textrm{OO}$, it is appreciable that the cooling causes
more pronounced shells around the chosen molecule, cf. figure~\ref{fig3}~(a). While at high temperatures
even the first coordination shell is not very well pronounced, it is
seen perfectly well at low temperatures, and the coordination number is around four as expected.
The derivative of the coordination number is more illustrative [figure~\ref{fig3}~(b)]. It shows the location of
the first, the second shell and even the third one, the latter being developed
at very low temperatures and serving as a manifestation of a spatial order that
develops in the system and involves a larger number of particles than the first neighbors.

To proceed, it is necessary to introduce the concept of H-bonds.
Energetic and geometric criteria, unavoidably involving certain degree of arbitrariness,
are used in the literature to interpret the structure of water and aqueous solutions.
A comprehensive analysis of several geometric definitions of hydrogen bonding, leading to different
values for the average number of H-bonds per molecule, has been performed in reference \cite{kumar}.

In this work, we use a single
geometric criterion, see e.g. Zhang et al. \cite{Zhang1} for two different water models.
Three conditions determine the existence of the H-bond, namely the distance $R_\textrm{OO}$ between
oxygen atoms belonging to two water molecules should not exceed the
threshold value $R_\textrm{OO}^\textrm{C}$; the distance $R_\textrm{OH}$ between the acceptor oxygen atom and the hydrogen
atom connected to the donor oxygen should not exceed the threshold value $R_\textrm{OH}^\textrm{C}$. Finally, the
angle O--H$\cdots$O should be smaller than the threshold value,
see e.g. references \cite{Padro1,Guardia1,Guardia2,Ioannis1,Ioannis2},
concerning the analysis of the structure of liquid methanol as well as of supercritical water.
If the coordinates of two molecules fulfill the above conditions, they are considered as H-bonded.
Zhang et al. proposed constant threshold values for the distances, 3.5~{\AA} and 2.45~{\AA},
for $R_\textrm{OO}^\textrm{C}$ and $R_\textrm{OH}^\textrm{C}$, respectively. However, the values used in this work are the
ones coming from the corresponding radial distribution functions minima for O--O and O--H
at each calculated temperature.
In contrast to \cite{Zhang1}, we used the threshold constant value for
the angle $\theta$, complementary to the
angle $\alpha$ between the vectors OH and $\bf{r}$$_\textrm{OH}$ (see figure~1 of reference \cite{kumar}),
that is equal to $150^{\circ}$. With this definition, we obtained the average number of
hydrogen bonds per water molecule similar to the value by Bak\'o et al. \cite{bako1} who used
the energetic criterion for bonding.

\begin{figure}[!t]
\begin{center}
\includegraphics[width=0.49\textwidth,clip]{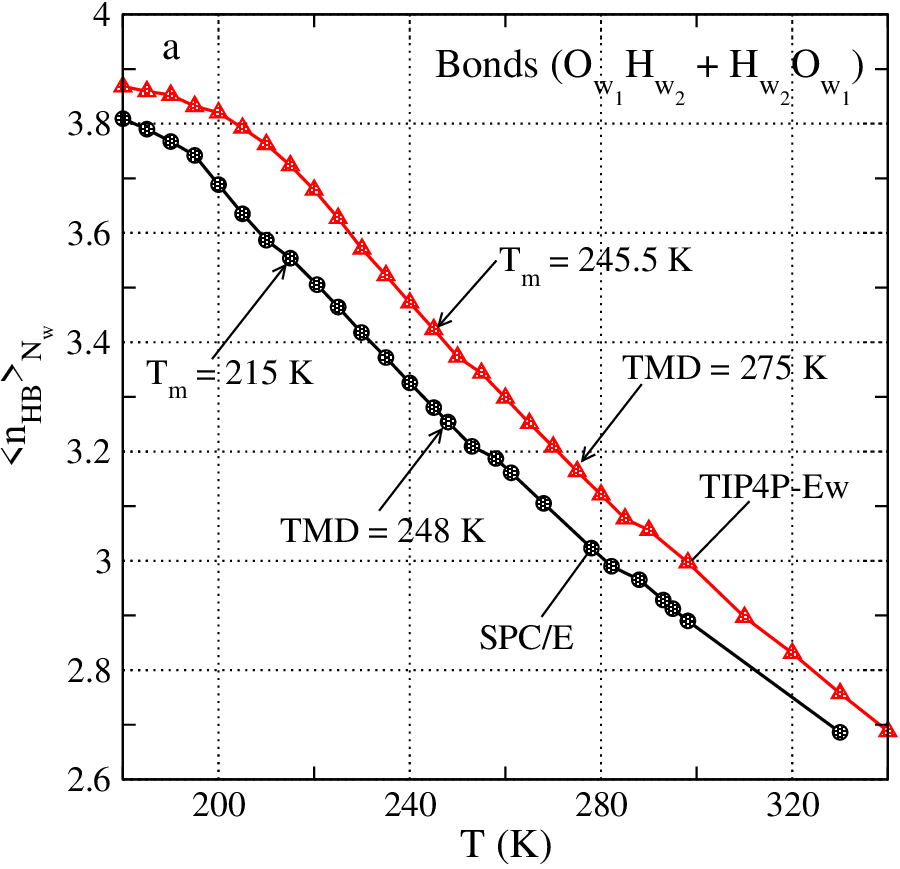}
\hspace{1mm}
\includegraphics[width=0.49\textwidth,clip]{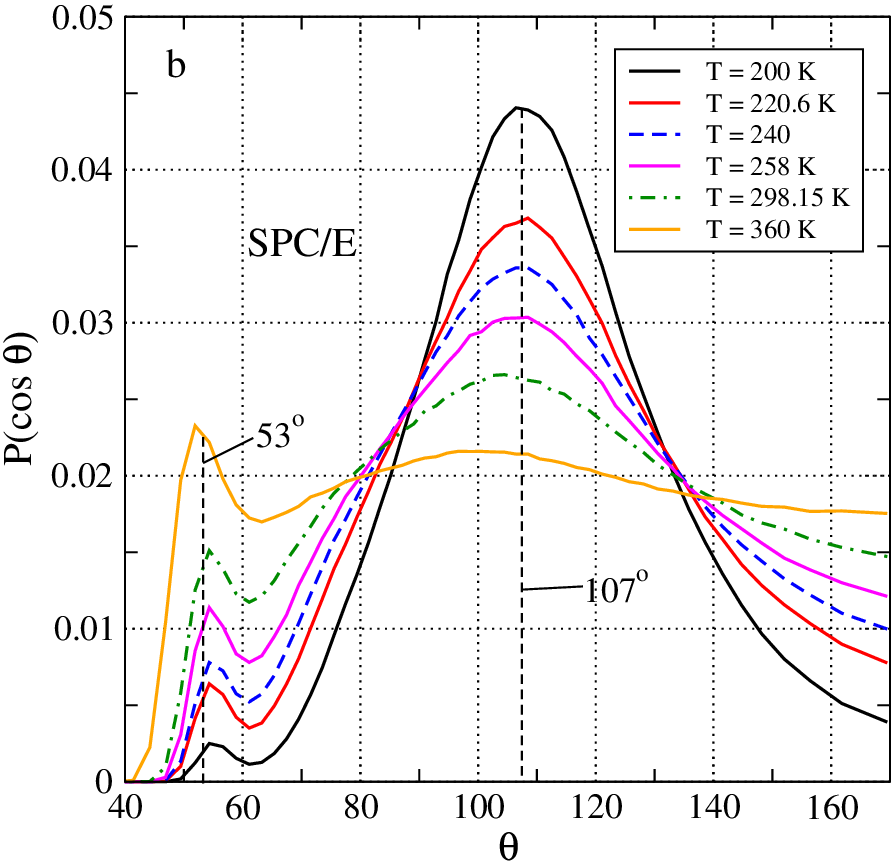}
\\[1ex]
\includegraphics[width=0.48\textwidth,clip]{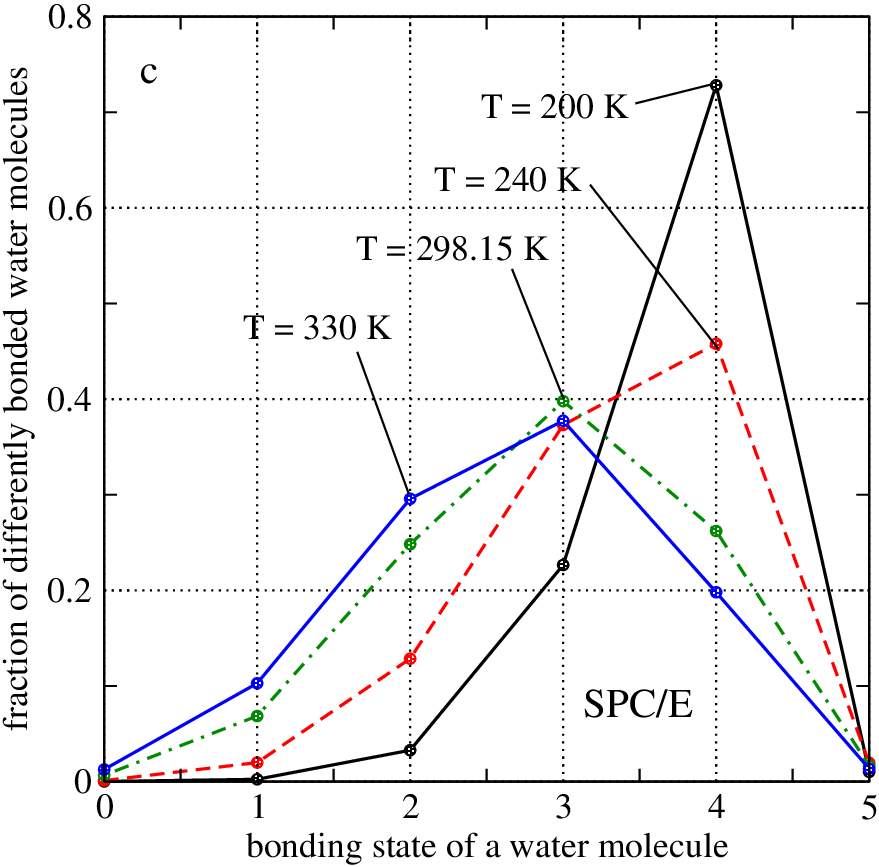}
\hspace{3mm}
\includegraphics[width=0.48\textwidth,clip]{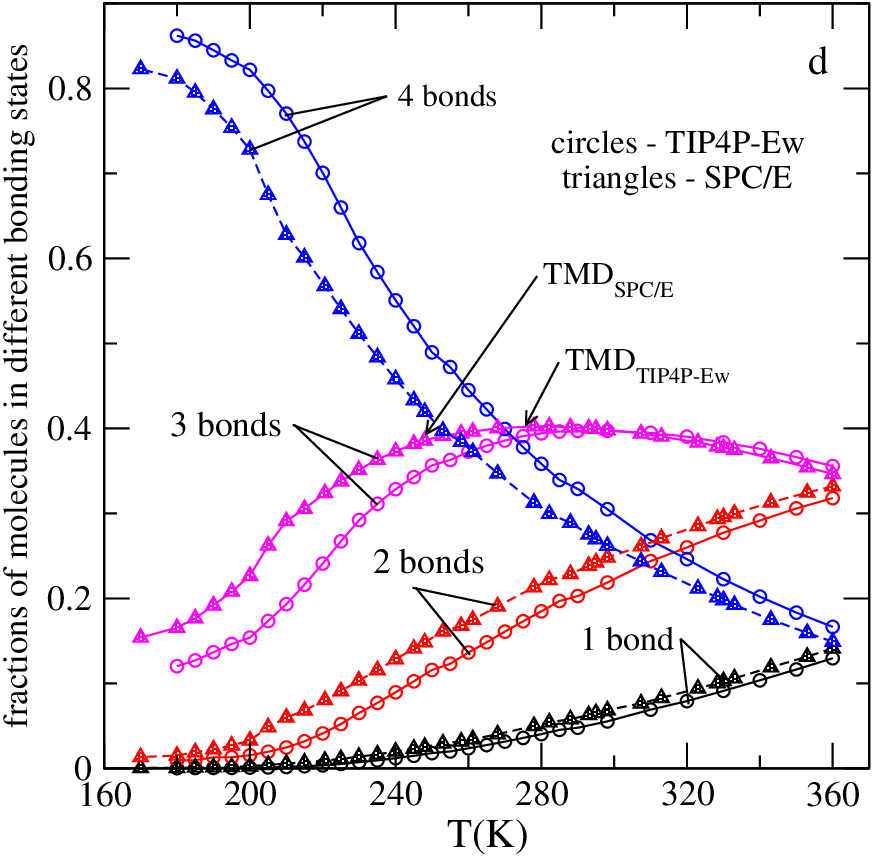}
\end{center}
\caption{(Color online) Temperature dependence of the average number of H-bonds
per water molecule (panel~a). Distribution function of the angles formed
by three oxygens on different molecules at different temperatures (panel b).
Fractions of water molecules in  different bonding
states on temperature (panel c). A comparison of fractions of differently
bonded molecules on temperature for the SPC/E and TIP4P-Ew models (panel d).}
\label{fig4}
\protect
\end{figure}

The temperature dependence of the number of H-bonds averaged upon simulation time and
normalized by the number of water molecules in the system, $\langle n_\textrm{HB}\rangle_{N_w}$,
following from our calculations for two water models, the SPC/E and TIP4P-Ew, is shown in figure~\ref{fig4}~(a).
This figure explains that a decreasing temperature results in a larger average number of
H-bonds. It can be seen that there is a change of the slope at $T \approx180$~K for the SPC/E model
where the transition into a glassy state is reported. On the other hand, the change of the slope
is observed at a higher temperature, around $T=200$~K, for the TIP4P-Ew model. However, this latter
model is characterized by a higher number of H-bonds per molecule in the entire temperature
interval studied. In both cases we see that the average number of bonds does not reach four,
or, in other words, even at low temperatures, perfect tetrahedral configuration is not attained.

The same conclusion can be reached by considering the ``bond-angle distribution function''
already studied by Bryk and Haymet \cite{haymet3}. It describes the probability to find
certain configurations involving three oxygens belonging to different water molecules.
The threshold distance is assumed to be at the first minimum of the pair distribution funcion,
$g_\textrm{OO}(r)$ at each temperature studied. The data were normalized by the number of
molecules that one finds in the shell limited by the threshold distance.
All the curves given in figure~\ref{fig4}~(b) are characterized by two
maxima. One of them at $\theta\approx53^\circ$ describes the configurations characteristic of a simple
liquid \cite{haymet2}. On the other hand, the second maximum at $\theta\approx107^\circ$ is close to the angle
$\theta\approx109.5^\circ$ describing a perfect tetrahedral configuration. At a high temperature, $T=360$~K,
a simple liquid type structure prevails. While the system is cooled, the first maximum decreases
and becomes very small at $T=200$~K, whereas the maximum describing the configurations close to tetrahedral
substantially grows in value. It is worth mentioning that both maxima slightly shift to higher angles
with a decreasing temperature. Still, even at low temperatures, the shape of the curves is not delta-like,
witnessing a certain degree of disorder in the system. Moreover, we do not observe any peculiarities
of this property either at TMD or at the temperature where the density is at minimum.

In order to obtain a better insight into the changes of the structure in terms of bonds with temperature,
we calculate the time-averages of fractions of differently bonded molecules. This representation
was already used  by Bak\'o et al. \cite{bako1} in their analysis of the
simulated structures of water-methanol mixtures at room temperature. Our results concerning
the SPC/E model at different temperatures are shown in figure~\ref{fig4}~(c). We observe that at a high temperature,
$T=330$~K, the fractions of molecules participating in two and three bonds are the largest.
Comparing this and room temperature, it can be seen that the fraction of molecules with four bonds
grows at the expense of a decreasing fraction of molecules with a single bond and two bonds. The fraction
of particles with three bonds remains practically unchanged. A further decrease of temperature down to
$T=240$~K (note that it is very close to the TMD) leads to a substantial increase of four-bonded
particles again at the expense of singly-bonded and doubly-bonded species.
The fraction of triple-bonded species remains intact. If the temperature falls
down to $200$~K, not only the fraction of 4-bonded particles grows substantially, but the fraction
of triple-bonded species exhibits a strong decrease in value. To summarize, the growth of density is
accompanied by the growth of the fraction of 4-bonded particles with almost constant fraction of
triple-bonded species. On the other hand, the descending density region, below the TMD, is characterized
by the growth of 4-bonded species with a simultaneous decrease of fractions of triple-bonded,
double-bonded and single bonded-molecules. Interestingly, the changes of the structure at even
lower temperature are very small (we do not show these curves to avoid overload of the figure).

The dependence of fractions of differently bonded molecules on temperature, in the entire temperature
range studied, is shown in figure~\ref{fig4}~(d). Here, we present our results for two different models, namely for the
SPC/E and TIP4P-Ew. The pattern is very similar in both cases. The TMD for each model obtained
from the calculations of density on temperature is located very close to the crossing points between
the curves describing the dependence of triple-bonded and 4-bonded molecules. The TMD values
are marked in the figure as well. It is worth mentioning
that the curves describing the fraction of triple-bonded molecules is not sensitive to the
temperature variable in the region between high temperatures and $T\approx240$~K. Moreover, the
values for this fraction coming from the calculations of two models in question
practically coincide in an ample interval of temperatures.
However, the difference between these two models is well observed due to their capability of forming
structures describing 4-bonded species.

In order to complete our discussion of the results given in all panels of figure~\ref{fig4}, we would like
to make the following comments. As mentioned above, the geometric definition of H-bonds
permits flexibility in the sense that one can attempt to change the values and the set of parameters
of H-bond definition. In this respect, the energetic definition is more restrictive. Besides
the choice of a set ($R_\textrm{OO}$, $r_\textrm{OH}$, $\theta$), we also tested the sets ($R_\textrm{OO}$, $r_\textrm{OH}$)
and ($R_\textrm{OO}$, $\beta$, cf. reference \cite{kumar} figure~1). These less restrictive geometric definitions
lead to the results qualitatively similar to the ones discussed above. Namely, the fraction of
quadruply-bonded species grows with a decreasing temperature whereas all other fractions decrease.
Unfortunately, we have not observed any characteristic crossing attributable to
the TMD with these definitions. Therefore, we hope to extend our study of the
models in question in the framework of energetic definition of H-bonding as well.

\begin{figure}[!t]
\begin{center}
\hspace{2mm}
\includegraphics[width=0.48\textwidth,clip]{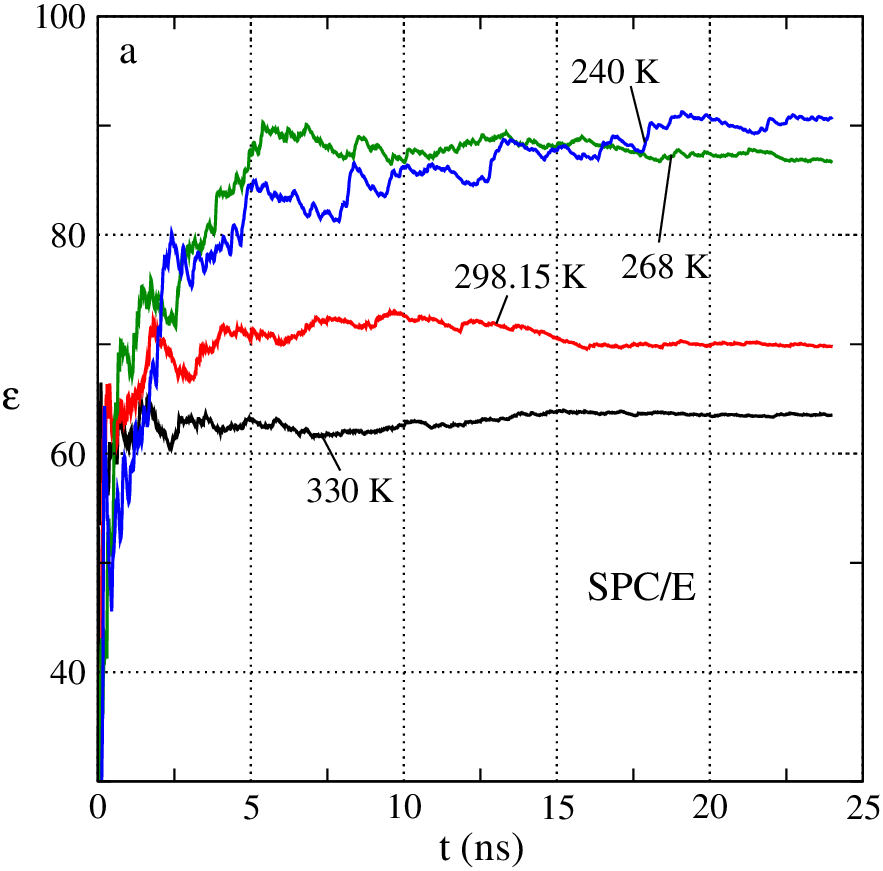}
\hfill
\includegraphics[width=0.48\textwidth,clip]{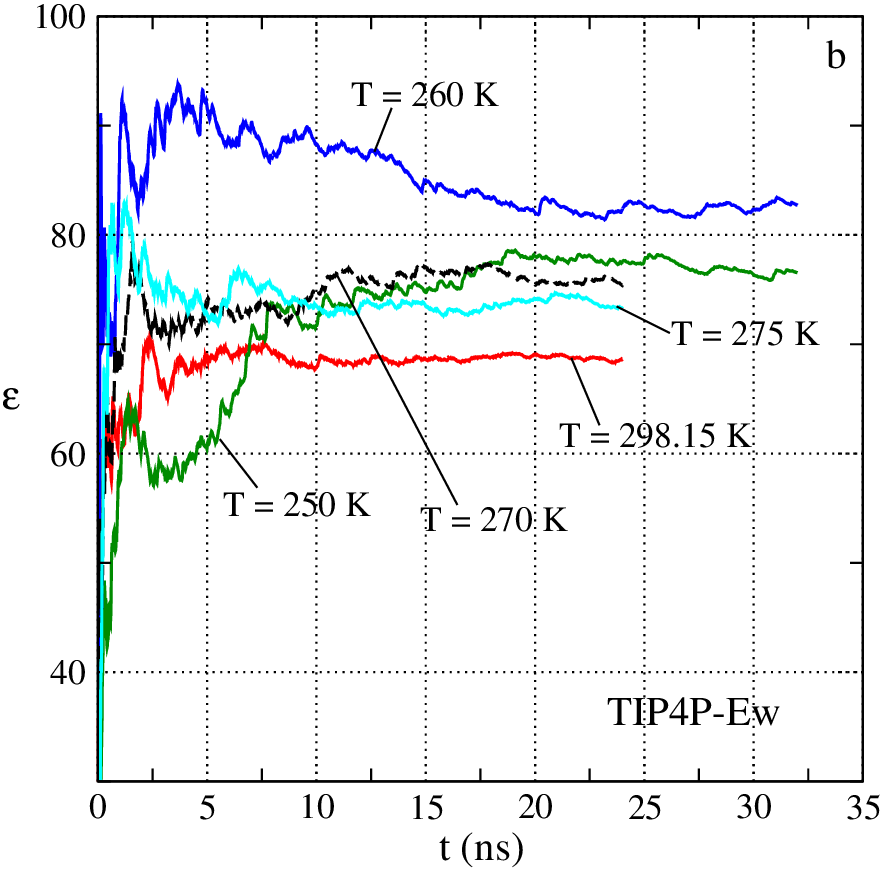}
\\[1ex]
\includegraphics[width=0.495\textwidth,clip]{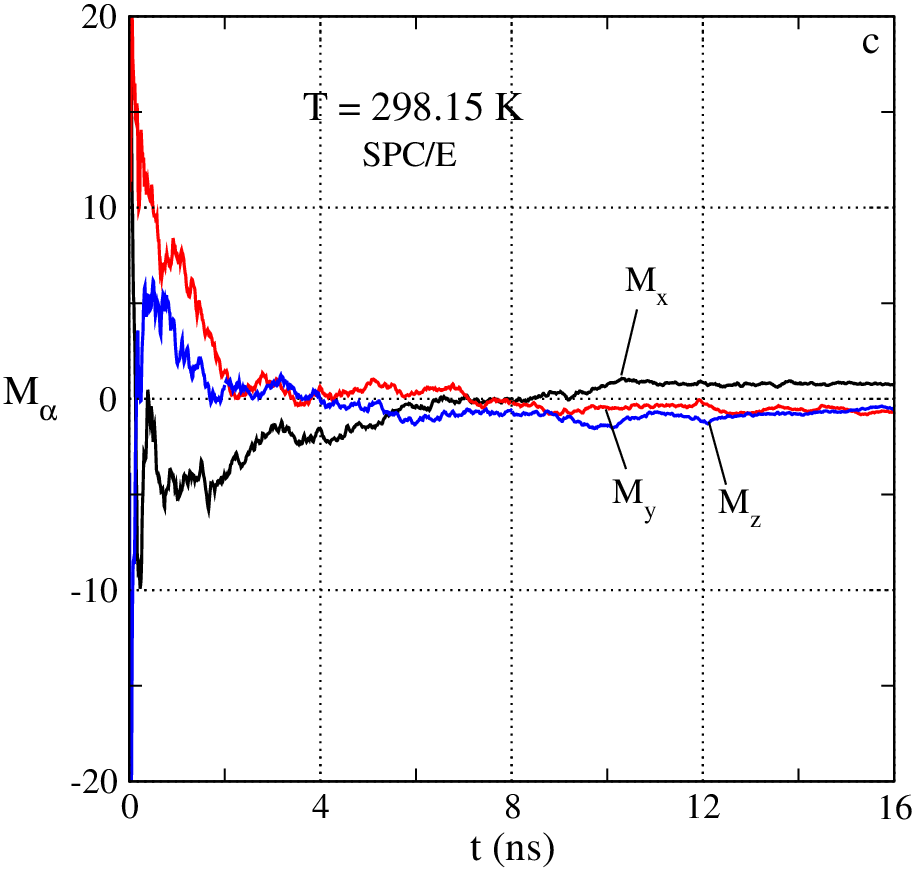}
\hfill
\includegraphics[width=0.47\textwidth,clip]{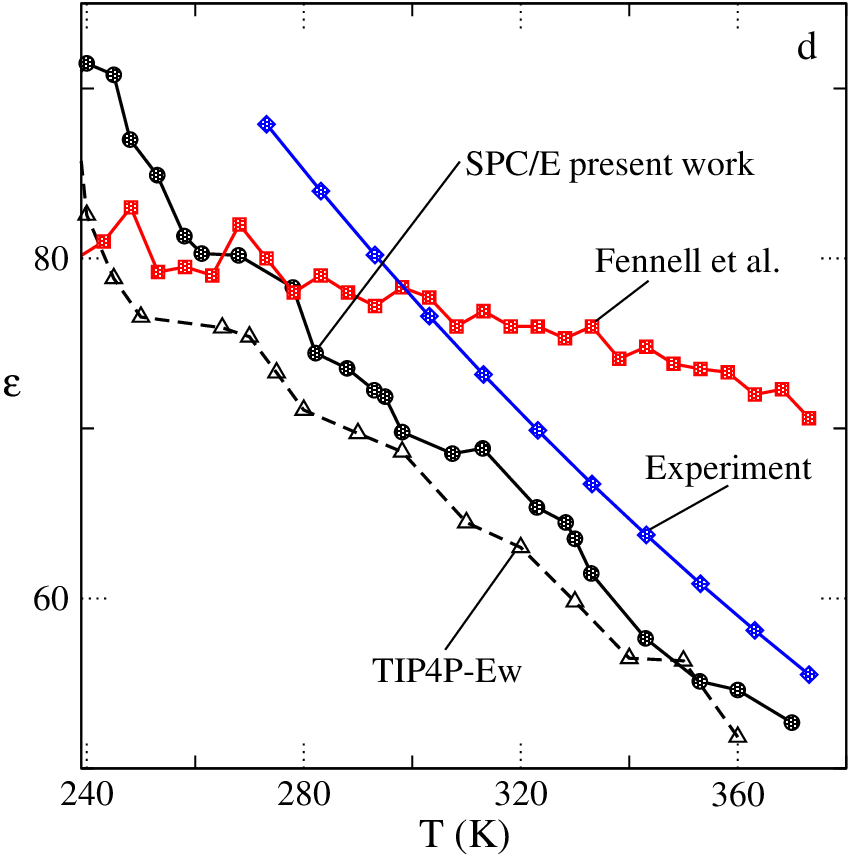}
\hspace{2mm}
\end{center}
\caption{(Color online) Time dependence of the dielectric constant of the SPC/E and TIP4P-Ew
water model at different temperatures from our NPT molecular dynamics simulations,
panels a and b, respectively. Time dependence of the Cartesian projections of
the total dipole moment of the system at room temperature for the SPC/E model
(panel c). Temperature dependence of the dielectric constant for the SPC/E and
TIP4P-Ew models from our calculations, from the fitting procedure by Fennell et al. \cite{dill}
and the experimental data \cite{handbook}.}\label{fig5}
\protect
\end{figure}

The properties described above refer to the structure of the systems in question mostly at short
inter-particle separation. On the other hand, the long-range, asympthotic behavior of correlations
between molecules possessing dipole moment is described by the dielectric constant, $\varepsilon$,
see e.g. \cite{myroslav}. In general terms, the calculation of the dielectric constant from
simulations is a difficult task \cite{alejandre,max,gereben}, especially at low temperatures. Several
factors can influence the result, such as e.g., the number of molecules, the type of thermostat and
barostat, precision of the long-range interactions summation.
Actually, very long runs are necessary to obtain reasonable estimates for this property,
as it is usually calculated from the time-average of the fluctuations of the total
dipole moment of the system \cite{martin} as follows:
\begin{equation}
\varepsilon=1+\frac{4\pi}{3k_\textrm{B}TV}\left(\langle{\bf M}^2\rangle-\langle{\bf M}\rangle^2\right),
\end{equation}
where $k_\textrm{B}$ is the Boltzmann constant and $V$ is the simulation cell volume.
Some of our results at different temperatures for the SPC/E model and for
TIP4P-Ew are shown in figure~\ref{fig5}~(a) and \ref{fig5}~(b), respectively.
We would like to recall here that all the data given in figure~\ref{fig5} were obtained for the
systems having two thousand particles. Practically all the runs lasted more than 20~ns.

Still, as it can be seen, there is room for experimenting in order to get better results
below the room temperature by extending the length of the runs or considering larger systems.
In addition, we plan to explore the
effects of changing the cutoff distance, of including the long-range corrections and the
precision of Ewald summation as well as to test other thermostats.
All the results for each model, however, refer to the temperatures above the respective melting point.
Still, there is one delimiting factor. It manifests itself in the polarization of the samples
at low temperatures. At room temperature, figure~\ref{fig5}~(c), the projections of the total dipole moment of
the system tend to zero at large times, whereas at lower temperatures, in the supercooled
regime and close to the formation of glassy states, the samples inevitably fall into configuration
with non-zero one or more Cartesian projections of $\bf M$. Then, the calculations of the dielectric constant
require different algorithms. A summarizing insight into our data for $\varepsilon(T)$ for two
models in question are given in figure~\ref{fig5}~(d). In general terms, the inclination of the temperature
dependence of the dielectric constant is correctly described by two models, compared to the
experimental results. The fitting of the dielectric constant at room temperature does not garantee
the correct temperature dependence, as we can see from the results of the modified model \cite{dill}.

To summarize this brief report, we have used
molecular dynamic simulations in the NPT ensemble to study the density anomaly of water
for the SPC/E and TIP4P-Ew models and to establish the relation between this anomaly
with the microscopic structure of the system. The structure is described in terms of
radial distribution functions, oxygen-oxygen coordination number and its derivative,
the average number of H-bonds per molecule, angular distributions and fractions of molecules
characterized by a different number of bonds.
We observed that at temperature of maximum density, the fractions of
molecules with three and with four bonds are almost the same, or in other words it is
a reasonably good structural estimator for the TMD. In addition, we presented our preliminary
results concerning the temperature dependence of the dielectric constant and compared them
with experimental data. While the slope of this dependence is well described, the values
are underestimated compared to the experimental results. It would be desirable in a future
work to extend our structural analysis to find estimates for the cooperative bonding of
molecules and relate them with the density and other anomalies of water. On the other hand,
it would be interesting to establish the relationship between the structural properties describing
the electric properties of the system and the dielectric constant.

\section*{Acknowledgements}
E.G. was supported by CONACyT of Mexico under Ph.D. scholarship. E.G. and O.P. are grateful to Dr.~T.~Patsahan for very helpful discussions and valuable comments. O.P. is grateful
to David Vazquez for technical support of this work.

\ukrainianpart


\title{Температурна залежність мікроскопічної структури і аномалія густини в моделях води SPC/E та TIP4P-Ew.
Результати моделювання методом  молекулярної динаміки}

\author{E. Галісія-Андрес\refaddr{label1}, Г. Домінгез\refaddr{label2},
 О. Пізіо\refaddr{label1}}
\addresses{
\addr{label1} Інститут хімії, Національний автономний університет Мексики,
Мехіко, Мексика
\addr{label2} Інститут дослідження матеріалів, Національний автономний університет Мексики,
Мехіко, Мексика
}

\makeukrtitle

\begin{abstract}
Ми дослідили температурні залежності мікроскопічної структури моделей води
SPC/E і TIP4P-Ew в термінах парних функцій розподілу, координаційних чисел, середнього числа водневих зв'язків,
розподілу зв'язаних станів однієї молекули, використовуючи метод молекулярної динаміки при постійногму тиску.
Еволюція структури поставлена у відповідність до залежності густини води від температури в області, що знаходиться між високими температурами і температурами, де система стає переохолодженою. Показано, що частка молекул з трьома і чотирма зв'язками визначає максимальну густину для обох моделей. Більш того, отримано і проаналізовано температурну залежність діелектричної сталої.
\keywords моделі води, аномалія густини, молекулярна динаміка
\end{abstract}

\end{document}